# Fast CNN-Based Object Tracking Using Localization Layers and Deep Features Interpolation

Al-Hussein A. El-Shafie
Faculty of Engineering
Cairo University
Giza, Egypt
elshafie_a@yahoo.com

Mohamed Zaki
Faculty of Engineering
Azhar University
Cairo, Egypt
azhar@eun.eg

S. E. D. Habib
Faculty of Engineering
Cairo University
Giza, Egypt
seraged@ieee.org

*Abstract*—Object trackers based on Convolution Neural Network (CNN) have achieved state-of-the-art performance on recent tracking benchmarks, while they suffer from slow computational speed. The high computational load arises from the extraction of the feature maps of the candidate and training patches in every video frame. The candidate and training patches are typically placed randomly around the previous target location and the estimated target location respectively. In this paper, we propose novel schemes to speed-up the processing of the CNN-based trackers. We input the whole region-of-interest once to the CNN to eliminate the redundant computations of the random candidate patches. In addition to classifying each candidate patch as an object or background, we adapt the CNN to classify the target location inside the object patches as a coarse localization step, and we employ bilinear interpolation for the CNN feature maps as a fine localization step. Moreover, bilinear interpolation is exploited to generate CNN feature maps of the training patches without actually forwarding the training patches through the network which achieves a significant reduction of the required computations. Our tracker does not rely on offline video training. It achieves competitive performance results on the OTB benchmark with 8x speed improvements compared to the equivalent tracker.

## I. INTRODUCTION

Visual object tracking is a classical problem in the computer vision domain where the location of the target is estimated in every video frame. The tracking research field continues to be active since long period because of the several variations imposed in the tracking process, like occlusion, changing appearance, illumination changes and cluttered background. It is challenging for a tracker to handle all these variations in a single framework. Therefore, numerous algorithms and schemes exist in literature aiming to tackle the tracking challenges and improve the overall tracing performance [1]-[3].

A typical tracking system consists of two main models, motion model and appearance model. The motion model is employed to predict the target location in the next frame like using Kalman filter [4] or particle filter [5] to model the target motion. The motion model can also be simple like constraining the search space to a small search window around the previous target location and assuming the target motion is small. On the other hand, the appearance model is used to represent the target and verify the predicted location of the target in every frame [6]. The appearance model can be classified to generative and discriminative methods. In generative methods, the tracking is performed by searching for the most similar region to the object [6]. In discriminative methods, a classifier is used to distinguish the object from the background. In general, the appearance model can be updated online to account for the target appearance variations during tracking.

Traditionally, tracking algorithms employed hand-crafted features like pixel intensity, color and Histogram of Oriented Gradients (HOG) [7] to represent the target in either generative or discriminative appearance models. Although hand-crafted features achieve satisfactory performance in constrained environments, they are not robust to severe appearance changes [8]. Deep learning using Convolution Neural Networks (CNN) has recently achieved a significant performance boost to various computer vision applications. Visual object tracking has been affected by this popular trend in order to overcome the tracking challenges and obtain better performance than that obtained by hand-crafted features. In pure CNN-based trackers, the appearance model is learned by a CNN and a classifier is used to label the image patch as an object or background. CNN-based trackers [8]-[10] achieved state-of-the-art performance in latest benchmarks [11], [12] even with simple motion models and no offline training. However, CNN-based trackers typically suffer from high computational loads because of the large number of the candidate patches and the training patches which are required in the tracking phase and the training phase respectively.

In this paper, we address the speed limitations of the CNN-based trackers. We adapt the CNN not only as a two-label classifier, object and background labeling, but also as a five-position classifier for the object position inside the candidate patch. This scheme allows achieving coarse object localization with less number of candidate patches. In addition, we exploit a bilinear interpolation scheme of the CNN feature maps already extracted in the coarse localization step for two purposes: first for the fine object localization, and second for the CNN feature extraction of the training patches. The computation of the bilinear interpolation is significantly less than that of extracting a new feature map which would speed-up the required processing time. Moreover, we did not perform offline training on any tracking dataset for our tracker.

This paper is organized as follows: Section II gives an overview of the CNN-based trackers and the speed bottlenecks in these systems. Our proposed schemes are presented in Section III. Section IV demonstrates the experimental results with the OTB benchmark, and finally, Section V concludes our work.

## II. OVERVIEW OF CNN-BASED TRACKERS

Following the huge success of deep CNNs in image classification [13], [14] and object detection applications [15], [16], many recent works in the object tracking domain have adopted deep CNNs and achieved state-of-the-art performance. There exists different use cases of CNNs in the tracking filed. References [17]-[19] employed CNNs with Discriminative Correlation Filters (DCF) where the regression models of these DCF-based trackers are trained by the feature maps extracted by the deep CNNs. References [20]-[22] adopted Siamese structure where two identical CNN branches are used to generate feature maps for two patches simultaneously either from the same frame or successive frames. The outputs from both branches are then correlated to localize the target. References [8]-[10], [23], [24] are pure CNN-based trackers where fully-connected layers are added after generating the feature maps to classify the input patches to object or background. A softmax layer is typically used at the end to score the candidate patches and opt the highest object score as the new target location. These pure CNN-based trackers achieved state-of-the-art performance in the latest benchmarks and we focus on this type of trackers in the rest of the paper.

Fig. 1 shows a typical CNN-based tracker. In each frame, candidate patches are generated with different translations and scales sampled from a Gaussian distribution. The mean of the Gaussian distribution is the previous location and scale of the target. The deep features are extracted by the convolution layers for each patch and scored by the fully connected (fc) layers.

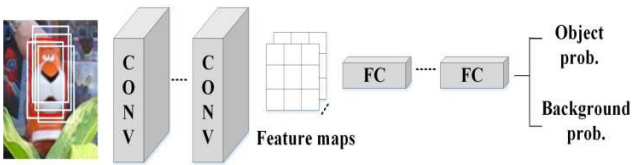

Fig. 1. A typical CNN-based tracker

For the training of the CNN-based trackers, transfer learning is typically exploited where the network parameters are initialized by another network pre-trained on large-scale classification dataset like ImageNet [25]. References [8]-[10] adopted offline training models to update the network parameters before tracking. It is difficult, however, to collect a large training dataset for visual tracking. Therefore, recent works [23], [24] dispensed with offline training steps and still achieved state-of-the-art performance. These techniques depend on increasing the number of iteration for the training in the initial frame because the target location is known and accurate. On the other hand, online training is necessary to cope with the potential appearance changes of the target. It is typical to update the parameters of the fully-connected layer only and keep those of the convolution layers fixed throughout the whole tracking process because the convolutional layers would have generic tracking information while the fully-connected layers would have target-background specific information. Short-term and long-term model updates proposed by [8] have been employed as well in other CNN-based trackers [9], [10], [23], [24]. Long-term updates are carried out at regular intervals, while short-term updates are carried out when the object score drops severely during tracking. The training data required for the online training is obtained every frame where deep features for positive and negative patches are generated and stored. The positive and negative patches have Intersection of Union (IoU) overlap with the estimated target location larger and less than certain thresholds respectively. When it is required to update the model, the stored positive and negative feature maps are sampled randomly to update the parameters.

The main computation steps in CNN-based trackers can be categorized into candidate evaluation, collecting training data and model update. The model update is performed at fixed intervals in the typical case and has less effect on the computation time compared to the candidate and training data processing. The CNN-based trackers mainly suffer from slow speed because of the computation in the convolutional layers to obtain the deep features for the candidate and training patches every frame. However, it can be noticed there would be a lot of computation redundancies because the candidate and training patches are generated randomly with large potential overlaps. Hence, we propose novel schemes in the next section to mitigate the redundant computations and speed-up the required processing time of the CNN-based trackers.

## III. PROPOSED CNN-BASED TRACKER

### A. Target localization

Although CNNs typically have local max-pooling layers to allow CNNs to be spatially-invariant to the input data, the intermediate feature maps are not actually invariant to large transformations of the input data [26]. Hence, we exploit this typical behavior of the network such that we do not only classify the patch into object or background, but also, classify the location of the object inside the patch. Having four classes as up, down, right and left to represent the target location inside the patch, we can localize the target with less number of candidates. In addition, we do not generate random candidate patches to cover the Region of Interest (ROI) as previous works but we generate fixed-spacing patches to cover the whole ROI as shown in Fig. 2. This scheme prevents the potential redundant computations in case of generating random patches and reduces the risk of missing the target. We propose also to forward the whole ROI through the convolution layers to save some redundant computations instead of forwarding each patch separately.

This idea is similar to what proposed in [16], [27] in the object detection field where the whole image is forwarded through the network instead of the proposal regions.

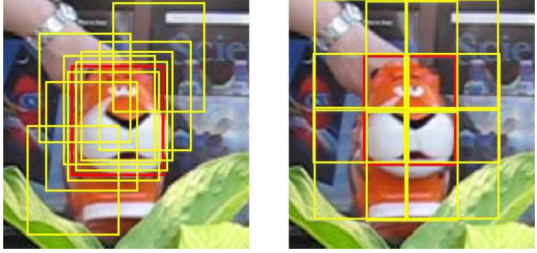

Fig. 2. Random patches and fixed-spacing patches

It is common in CNN-based trackers that the target localization is carried out by taking the mean location of the candidate patches with top object scores, while, in our scheme, the patches which are classified as objects are first moved based on the localization network. The patch with the highest overlap with the other object patches is selected as an input to a fine localization step where we utilize bilinear interpolation of the feature maps. Bilinear interpolation was first proposed by [26] for the implementation of a spatial transformer network and it was then employed by [28] in a ROI align scheme for object detection applications. Let's assume the target is represented by 3x3xd feature maps as shown in Fig. 3 (a), where d is the feature depth, and we extract feature maps for a region larger than the target size such that we get 5x5xd feature maps as shown in Fig. 3 (b). We would have nine 3x3 grid in total. Each 3x3 grid is displaced by dx and/or dy from its neighbors. The value of dx and dy depends on the network structure. Accordingly, we can obtain the feature maps of all image patches which have displacements ranging from 0 to dx or dy measured from the center by bilinear interpolation without forwarding the image patches through the convolution layers. Any point value is calculated by bilinear interpolation from the four nearby points in the feature maps such as point * in Fig. 3 (c).

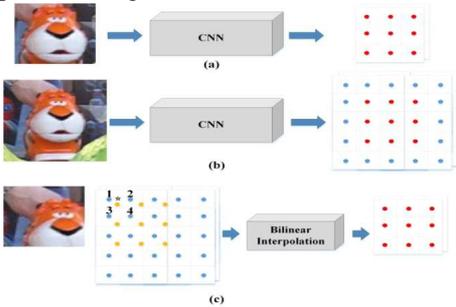

Fig. 3. Interpolation of feature maps

### B. Network training

We reuse the feature maps obtained in the localization phase to extract the feature maps of the positive and negative training patches by applying bilinear interpolation. The positive patches are actually sub-divided into localization patches. Although we add more classification classes in the network for localization, the required computation does not increase much because the localization patches are not forwarded through the whole convolutional layers and bilinear interpolation is exploited instead.

### C. Scale variation

Reference [8] handled scale variation by generating training and candidate patches with random scales drawn from a Gaussian distribution and forwarding these patches through the whole network to obtain the feature maps. However in our proposed scheme, we extract feature maps for three fixed scales only {1, max_scale_up, max_scale_down}. We then obtain the feature map of any required scale in that range, for either a candidate or training patch, by applying linear interpolation on two scales. Hence, instead of forwarding the images patches generated randomly in spatial and scale domains through the convolution layers, we extract feature maps for a larger image patch at three fixed scales. We perform bilinear interpolation to obtain the feature map at the required displacement and perform linear interpolation to obtain the feature map at the required scale. Fig. 4 illustrates our scheme of obtaining feature maps of image patches at different scales.

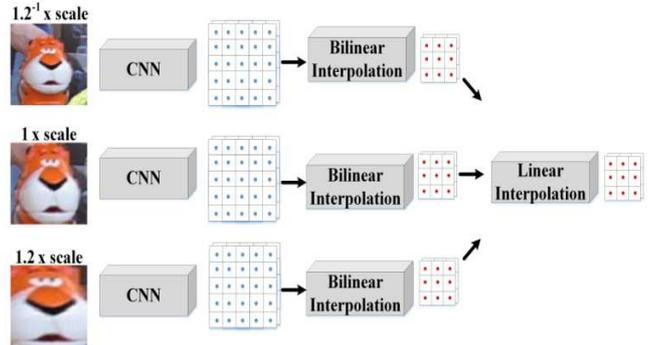

Fig. 4. Interpolation of fixed-scale feature maps

## IV. IMPLEMENATION DETAILS

### A. Netwrok stucture

We start with the MDNet_N implementation as a baseline for our work. MDNet_N is the same as MDNet [8] but without both offline training and bounding box regression. The parameters of the convolutional layers (conv1-3) are initialized by the VGG-M [29] model and the fully connected layers are initialized by random values. In [8], the object size (h×w) is cropped and padded to the network input size which is 107×107 such that this fixed size, 107×107, would be equivalent to an image patch of (107÷75)×(h×w). The spatial size of the feature maps generated from conv3 is 3×3 for a network input of 107×107. Our network shown in Fig. 5 is similar to MDNet but we add fc7-9 as a localization network and allow different input sizes to get feature maps of sizes 3×3, 5×5, 7×7 … etc when needed. The localization layer classifies the positive patches into five classes based on the position of the object inside the patch (up, down, right, left and middle).

*B. Initial frame training*

In order to generate training data for the object and localization layers, in the initial frame, we generate feature maps for an input of size 139×139 at three fixed scales: 1, 1.2 and 1.2-1. The output from conv3 would be 5x5 at three fixed scales. The initial object is actually represented by the inner 3x3 feature maps at scale 1. Accordingly, we can exploit bilinear interpolation and generate feature maps for any patch with a displacement ranging from 0 to a (16÷75 × w) and (16÷75 × h) in the x and y direction respectively and with different scales ranging from 1.2-1 to 1.2. The object training samples are generated from a Gaussian distribution in the same way as MDNet_N so that the IoU with the initial target location is larger than 0.7. The localization training samples are generated from five Gaussian distributions equivalent to each localization class and the IoU should be larger than 0.7 as well. In order to generate training data for the background, in the initial frame, we divide the background training data into two types, close and far samples. The close samples are the samples which are close to the initial target location, and hence, we can apply the same interpolation scheme used for the object and localization training samples. For the far background samples, we generate feature maps as normal by forwarding the samples through all the convolutional layers. All the background training samples should have IoU less than 0.5 with the initial target. Our network is trained by Stochastic Gradient Descent (SDG) with mini-batch size of 128 and 65 for fc4-6 and fc7-9 respectively and 90 iterations in the initial frame.

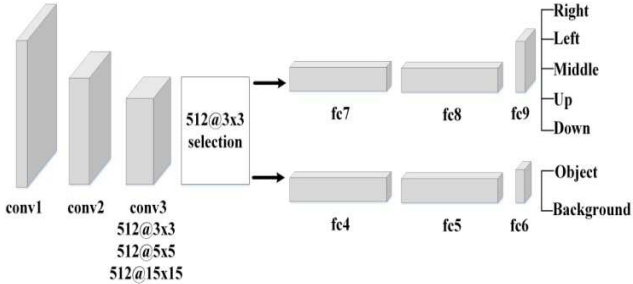

Fig. 5. Proposed network structure

*C. Object tracking*

We forward the whole ROI of size (4w×4h) which is centered on the previous target location through the convolution layers. We crop this ROI to 299×299 before entering the network and obtain a 15×15 conv3 feature map accordingly. As the object is represented by a 3x3 conv3 feature map, we would obtain 169 feature maps of spatial size 3x3. These 169 feature maps represent image patches displaced from the center by [k × (16÷75) × w] and [k × (16÷75) × h] in the x and y direction respectively, where k is an integer in [-5:5]. The object score of each 3x3 feature map is checked and if it is larger than 0.5, the new location of the equivalent patch will be obtained based on the localization network. The patch with the highest overlap with other object patches is chosen for the next fine localization step.

In the fine localization step, we need to find a finer location and a newer scale of the object. We calculate new 5x5 feature maps centered on the coarse location at two scales, 1.05 and 1.05-1. We generate 100 fine samples displaced with fixed values in the x and y direction and with different scales drawn from a Gaussian distribution. The feature maps of these fine samples are calculated by bilinear interpolation in the spatial and scale domains. Then, we check the object score of the fine samples and average the three samples with the highest object score.

*D. Online network update*

We adopt long-term and short-term network update schemes as proposed in [8]. Long-term is carried-out every 10 frames using the training samples collected, while, a short-term update is carried out when the object score is less than 0.5. We generate training samples for the object, background and localization layers each frame if the object score obtained in that frame is larger than 0.5 similar to [8]. However, we reuse the feature maps generated in the tracking stage to obtain the feature maps of the training samples by bilinear interpolation in the spatial and scale domains. In addition, we employ hard minibatch mining for the negative training samples similar to [8] where 96 negative samples of the highest positive score are selected out of 1024 negative samples. The number of training samples for the object, localization and background layers is 30, 30 and 100 samples respectively each frame and 10 iterations are adapted for the online update.

*E. Experimental results*

We evaluate our tracker on the Object Tracking Benchmark (OTB-100) [11] which contains 100 fully annotated videos. Our tracker is implemented in Matlab using MatConvNet and runs on an Intel i7-3520M CPU system. We ran MDNET_N on the same system as a reference system. The tracking performance is measured by performing One-Pass Evaluation (OPE) on two metrics, center location error and IoU between the estimated target location and the ground truth. Fig. 6 shows the precision plot and the success plot of our tracker on 100 video frames [11] compared with MDNET_N. The precision plot shows the percentage of frames whose estimated target location is within the error threshold (x-axis) of the ground truth, while the success plot shows the percentage of frames whose IoU is larger than the overlap threshold (x-axis). The legend values in the precision and the success plots are the precision score in case error threshold = 20 pixel and the area under the curve (AUC) of the success plot respectively. It can be seen from Fig. 6 that our tracker which is based on Interpolation and Localization Network (ILNET) has the same AUC as MDNET_N and slightly lower precision.

Fig. 7 demonstrates the effectiveness of our tracker to handle all kinds of tracking challenges. It can be seen that our tracker achieves almost the same or better performance compared to the baseline tracker, MDNET_N. Table I shows the breakdown of the processing time savings achieved by our tracker. Both the tracking and the training speeds have increased despite adding a localization network and increasing the number of training iterations. This speed-up achievement is due to bilinear interpolation on the feature maps and by using fixed-spaced candidates.

TABLE I AVERAGE COMPUTATION TIME IN SECONDS PER FRAME

|  | MDNET-N[*] [8] | Our work (ILNET) | Speed-up factor |
|---|---|---|---|
| Candidate processing | 3.4 | 0.36 | 9.4x |
| Training processing | 3.3 | 0.21 | 15.7x |
| Network update (@10th frame for long-term) | 2.3 | 2.3 | 1x |
| First frame training | 90 | 52 | 1.72x |
| Frame processing without first frame | 7 | 0.8 | 8.8x |

[*]MDNET_N:MDNET without offline training and bounding box regression

During the preparation of our paper, another recent work [30] was published which adopts a bilinear interpolation scheme similar to the one used in our paper. However, our work has three distinct features: first, we do not offline-train our network on any video dataset. Secondly, we input the ROI only to the convolutional layers, while in [30], almost the whole image is forwarded. Finally, we add a localization network in order to test fixed-spaced candidates instead of random candidates which would allow us to cover the whole ROI and reduce the probability of missing the target.

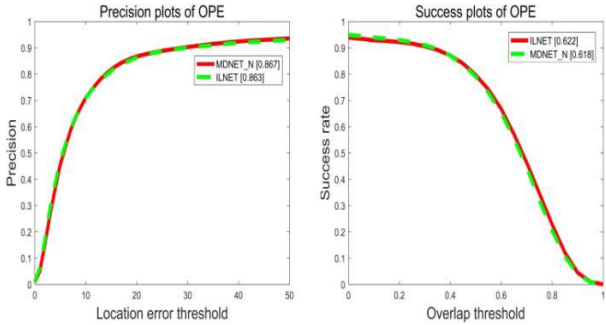
Fig. 6. Precision and success plots on OTB100

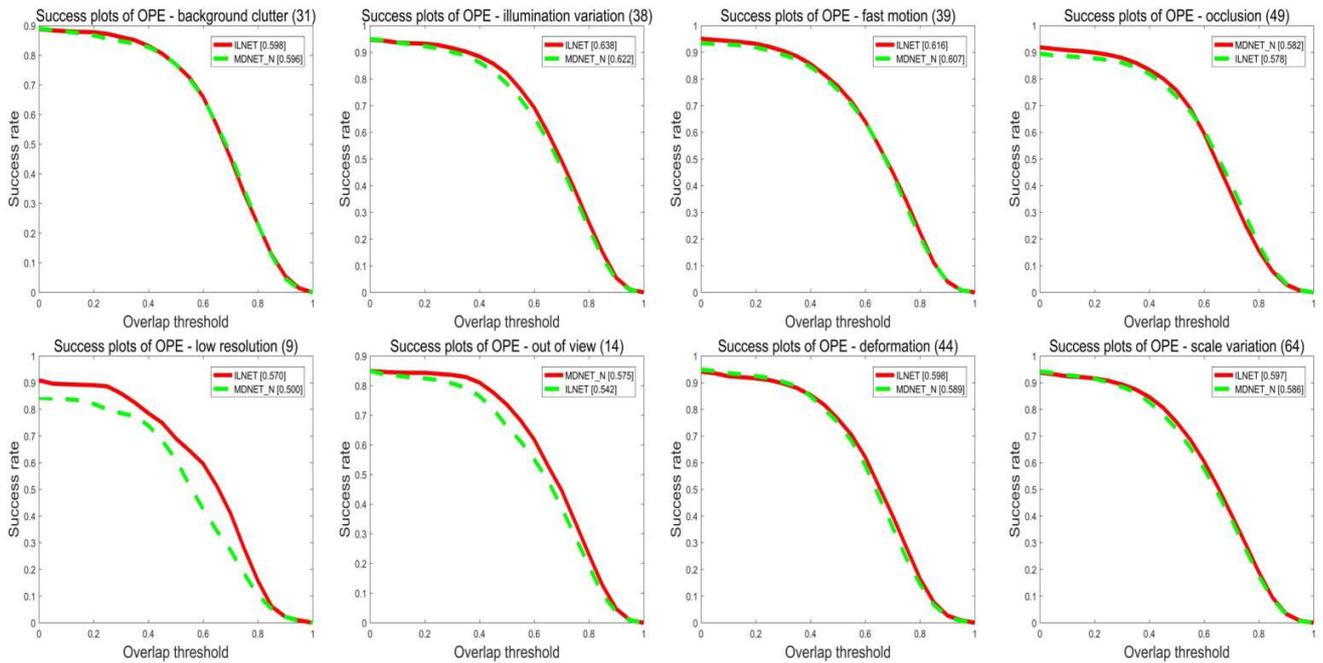
Fig. 7. The success plot of eight challenge attributes

## V. CONCLUSION

We present a fast CNN-based object tracker where we address the speed issues of the typical CNN-based trackers. The main computation overhead in CNN-based trackers arises from forwarding the candidate and training patches through all the convolutional layers to extract the feature maps of the image patches. In this paper, we reduce the redundant computation of feature maps for the candidate and training patches by employing bilinear interpolation. In addition, we add a localization network and forward the whole ROI once through the convolutional layers which allows us to cover the whole ROI with fixed-spaced candidates instead of random candidates. Consequently, the required computation is reduced. Our network is just initialized by the VGG-M network pertained on the ImageNet dataset, and it is not offline-trained on any video dataset. Our proposed CNN-based tracker employs simple tracking and training stages which would facilitate the embedded implementation including the hardware implementation. Our tracker achieved the same performance as the baseline tracker on the OTB dataset while featuring 8x speed-up improvement.